\documentclass{article}

\input{tcilatex}

\begin{document}

\textbf{QUATERNION DYNAMICS OF THE BRAIN}

\bigskip

E. A. Novikov

\bigskip

Institute for Nonlinear Science, University of California - San Diego, La
Jolla, CA 92093 - 0402

\bigskip

A nonlinear dynamical modeling of interaction between automatic and
conscious processes in the brain is described. Effects of sensations,
emotions and reflections on the electromagnetic activity of the brain are
represented in terms of quaternion field.

\bigskip

In previous paper [1] an approach to nonlinear dynamical modeling of
interaction between automatic (A) and conscious (C) processes in the brain
was presented. The idea is to use complex field with real and imaginary
components representing A- and C-processes. The interaction is due to the
nonlinearity of the system. This approach was illustrated on the nonlinear
equation for the current density in the cortex. The nonlinearity is
determined by the sigmoidal firing rate of neurons. More general approaches
were also indicated [1]. In this letter the quaternion (Q) approach is
described.

Subjective C-experiences can be divided into three major groups: sensations
(S), emotions (E) and reflections (R). Note, that subjective S should be
distinguished from the automatic sensory input into the neuron system of the
brain. The effects of S, E and R on the electrochemical activity of the
brain, generally, can be different. These effects should be distinguished in
designing experiments and in C-modeling. The natural generalization of the
approach [1] is to use the Q-field for representation of S-, E- and
R-effects.

Let us illustrate this on the model equation for the average (spatially
uniform) current density $\alpha (t)$ perpendicular to the cortical surface
[1]:

\begin{equation}
\frac{\partial \alpha }{\partial t}+k\alpha =f(\alpha +\sigma )  \tag{(1)}
\end{equation}%
Here $k$ is the relaxation coefficient, $\sigma (t)$ is the average sensory
input and $f$ \ represents the sigmoidal firing rate of neurons, for
example, $\ f(\alpha )=\tanh (\alpha )$. For the case of spatially
nonuniform $\alpha (t,\mathbf{x})$ and $\sigma (t,\mathbf{x})$ we can use a
more general equation, which include typical propagation velocity of signals 
$v$ [1]:

\begin{equation}
\frac{\partial ^{2}\alpha }{\partial t^{2}}+(k+m)\frac{\partial \alpha }{%
\partial t}+(km-v^{2}\Delta )\alpha =(m+\frac{\partial }{\partial t}%
)f(\alpha +\sigma )  \tag{(2)}
\end{equation}%
Here $m$ is an additional parameter and $\Delta $ is the two-dimensional
spatial Laplacian. This type of equations are used for interpretation of the
EEG and MEG spatial patterns (see recent paper [2] and references therein).
In this context we have parameters: $k\sim m\sim v/l$, where $l$ is the
connectivity scale.

Let us now introduce the Q-field:

\begin{equation}
q=\alpha +i_{p}\psi _{p}  \tag{(3)}
\end{equation}%
Here components $\psi _{p}$ represent (S, E, R)-effects and summation is
assumed on repeated subscripts from $1$ to $3$. The quaternion imaginary
units $i_{p}$ satisfy conditions:

\begin{equation}
i_{p}i_{q}=\varepsilon _{pqr}i_{r}-\delta _{pq}  \tag{(4)}
\end{equation}%
where $\varepsilon _{pqr}$ is the unit antisymmetric tensor and $\delta
_{pq} $ is the unit tensor. Formula (4) is a compact form of conditions: $%
i_{1}^{2}=i_{2}^{2}=i_{3}^{2}=-1,$ $i_{1}i_{2}=-i_{2}i_{1}=i_{3},$ $%
i_{2}i_{3}=-i_{3}i_{2}=i_{1},$ $i_{3}i_{1}=-i_{1}i_{3}=i_{2}$. Conjugate
Q-field is defined by: $\bar{q}=\alpha -i_{p}\psi _{p}$. From (3) and (4) it
follow that the components and modulus of Q-field can be obtained by
formulas:

\begin{equation}
\alpha =\func{Re}\{q\}=\frac{1}{2}(q+\bar{q}),\text{ }\psi _{p}=\func{Im}%
_{p}\{q\}=-\func{Re}\{qi_{p}\},\text{ }\alpha ^{2}+\psi _{p}^{2}=\mid q\mid
^{2}=q\bar{q}  \tag{(5)}
\end{equation}%
The Q-multiplication, used in (5), is determined by (4). The Q-inversion is
defined by:

\begin{equation}
\frac{1}{q}=\frac{\bar{q}}{\mid q\mid ^{2}}  \tag{(6)}
\end{equation}%
Note, that the only finite-dimentional division algebras over the real
number field are the real numbers, the complex numbers and the Q-field.
Functions of Q-field are defined similarly to functions of complex fields.
For example:

\begin{equation}
\exp (q)=\exp (\alpha )[\cos (\psi )+j\sin (\psi )\text{ }],\text{ }\psi
^{2}\equiv \psi _{p}^{2},\text{ }j\equiv i_{p}\psi _{p}\psi ^{-1},\text{ }%
j^{2}=-1  \tag{(7)}
\end{equation}%
Using (4) - (7), we can define many Q-functions, particularly:

\begin{equation}
\tanh (q)=\frac{\exp (q)-\exp (-q)}{\exp (q)+\exp (-q)}=\frac{\exp (2\alpha
)-\exp (-2\alpha )+2j\sin (2\psi )}{\exp (2\alpha )+\exp (-2\alpha )+2\cos
(2\psi )}  \tag{(8)}
\end{equation}

Returning to (1), we now substitute into this equation Q-field (3) instead
of $\alpha $. It gives four equations:

\begin{equation}
\frac{\partial \alpha }{\partial t}+k\alpha =\func{Re}\{f(\alpha +\sigma
+i_{q}\psi _{q})\}  \tag{(9)}
\end{equation}

\begin{equation}
\frac{\partial \psi _{p}}{\partial t}+k\psi _{p}=\func{Im}_{p}\{f(\alpha
+\sigma +i_{q}\psi _{q})\},\text{ }(p=1,2,3)  \tag{(10)}
\end{equation}%
These equations are coupled because $f$ is nonlinear. Thus, we got
Q-modeling of the C-effects. The same scheme can be applied to equation (2)
and to any nonlinear model equation. For $f(\alpha )=\tanh (\alpha )$,
formulas (7) and (8), with shift $\alpha \Longrightarrow \alpha +\sigma $,
give explicit form for the nonlinear terms in equations (9) and (10). Note,
that so-called extra-sensory effects (if they exist) can be included in this
approach by assuming that $\sigma $ has imaginary components. When $\alpha $
and $\sigma $ are relatively small (in a dream or in a state of deep
meditation), asymptotically (10) gives closed system of equations for $\psi
_{p}$. The nonlinear terms $\func{Im}_{p}\{f(i_{q}\psi _{q})\}$ are quite
different from the nonlinear term $f(\alpha )$ for the A-process. This may
explain peculiar dynamics of dreams and deep meditations, not only in
content, but also in intensity.

The obtained description of the C-effects can serve for designing
experiments to test the modeling. The chosen sequence of S-E-R-effects seems
to be natural for the Q-modeling.

It will be interesting to apply Q-modeling to the first principles,
particularly, to the problem of unification of the four major forces
(gravitational, electromagnetic, week and strong). The fundamental obstacle
to the unification is that gravitation in its nature is quite different from
the three other forces. It resists quantization with complex amplitudes,
which is the natural description for the other forces. Perhaps, in the
Q-description, gravitation can be an analog of real (automatic) process,
while other three forces are more " vibrating" - analog of S-E-R-effects in
the "Brain - Universe". In the Q-description "ghosts" [3,4] with the
negative energy may have natural explanation.

\bigskip

\bigskip

\bigskip

\bigskip

\textbf{References}

\bigskip

[1] E. A. Novikov, arXiv:nlin.PS/0309043 (2003)

[2] V. K. Jirsa, K. J. Jantzen, A. Fuchs, and J. A. S. Kelso, IEEE Trans. on
Medical Imaging, \textbf{21}(5), 497 (2002)

[3] C. V. Johnson, \emph{D-Branes}, Cambridge Univ. Press, 2003

[4] Y. Fujii and K. Maeda, \emph{The Scalar-Tensor Theory of Gravitation},
Cambridge Univ. Press, 2003

\end{document}